\documentclass[10pt,twocolumn,twoside]{IEEEtran}

\usepackage{graphicx,epsfig}
\usepackage{cite}
\usepackage{stfloats}
\usepackage{amsmath}
\usepackage{amsfonts,helvet}

\newtheorem{theorem}{Theorem}

\newtheorem{algo}[theorem]{Algorithm}

\newtheorem{proposition}[theorem]{Proposition}

\newcounter{proposition}
\begin{document}

\title{Network Massive MIMO for Cell-Boundary Users: From a Precoding Normalization Perspective}


\author{\authorblockN{Changwoo Lee,$^{\ast}$ Chan-Byoung Chae,$^{\ast}$  Taehyung Kim,$^{\dagger}$ Sooyong Choi,$^{\dagger}$} and Juho Lee$^{\ddagger}$\\
\authorblockA{$^{\ast}$School of Integrated Technology\\
$^{\dagger}$School of Electrical and Electronic Engineering\\
Yonsei University, Korea\\
Email: \{lcwyu7, cbchae, khotdog, csyong\}@yonsei.ac.kr\\
$^{\ddagger}$DMC R\&D Center, Samsung Electronics, Suwon, Korea\\
Email: juho95.lee@samsung.com\\
\thanks{This work was in part supported by Samsung Electronics, the KCC (Korea Communications Commission), Korea, under the R\&D program supervised by the KCA (Korea Communications Agency) (KCA-2012-1291101106) and the Ministry of Knowledge Economy under the IT Consilience Creative Program (NIPA-2012-H0201-12-1001).}}}



\maketitle \setcounter{page}{1} 
%
%
%


\begin{abstract}
In this paper, we propose network massive multiple-input multiple-output (MIMO) systems, where three radio units (RUs) connected via one digital unit (DU) support multiple user equipments (UEs) at a cell-boundary through the same radio resource, i.e., the same frequency/time band. For precoding designs, zero-forcing (ZF) and matched filter (MF) with vector or matrix normalization are considered. We also derive the formulae of the lower and upper bounds of the achievable sum rate for each precoding. Based on our analytical results, we observe that vector normalization is better for ZF while matrix normalization is better for MF. Given antenna configurations, we also derive the optimal switching point as a function of the number of active users in a network. Numerical simulations confirm our analytical results.

\end{abstract}


\section{Introduction}
Multiple-input multiple-output (MIMO) wireless communication techniques have evolved from single-user to multiple-user systems \cite{Chae_SPMag_07}. To approach the capacity of the MIMO broadcast channel, the authors in \cite{SpencerSwindle04, Chae_JSAC07} proposed simple zero-forcing (ZF) based-linear algorithms, where the transmitter and the receivers are equipped with multiple antennas. The optimality of the linear algorithm was intensively investigated in \cite{Chae_SPL09} with an assumption of an infinite number of antennas at the receiver. The authors in \cite{Chae_SPL09} proved that a simple linear beamforming (coordinated beamforming in the paper) asymptotically approaches the sum capacity achieved by dirty paper coding (DPC). 

Recently, massive MIMO (a.k.a. large-scale MIMO) has been proposed to further maximize network capacity and to conserve energy~\cite{Marzetta_06, Ngo_TCOM12, Marzetta_WC10}. In~\cite{Marzetta_06}, the authors showed that a single-cell system with an unlimited number of antennas at the transmitter is always advantageous. The authors in~\cite{Ngo_TCOM12} investigated the energy and spectral efficiency for massive MIMO systems for a single-cell environment. It was shown, however, that the capacity gains obtained by multiuser MIMO processing degrade severely in multi-cell environments. To further maximize the network capacity, several network MIMO algorithms with multiple receive antennas have been proposed~\cite{Chae_NCBF08, Chae_IACBF09}. These systems assume, however, that the network supports maximum of three users through a relatively small number of transmit antennas.\footnote{Note that more than three users can be supported if there is a common message, i.e., for a clustered broadcast channel.} 

Massive MIMO systems in multi-cell environments were also studied in~\cite{Marzetta_WC10, Jubin_WC11}. As investigated in~\cite{Marzetta_WC10, Jubin_WC11}, multi-cell massive MIMO has some critical issues, such as pilot contamination, that have to be resolved before it can be applied in practice. In~\cite{Jubin_WC11}, the authors showed theoretically and numerically the impact of pilot contamination, proposing a multi-cell minimum mean square error (MMSE) based precoding algorithm to reduce both intra- and inter-cell interference. In~\cite{Jubin_WC11}, matched filter (MF) precoding was used. Inter-user interference is eventually eliminated once the transmitter has large enough number of antennas. The assumption of an infinite number of antennas at the transmitter, however, is not, in practice, really feasible. This issue was studied in \cite{Jakob, Hoon11}. The author in \cite{Hoon11} concluded that the proposed architecture achieves the same spectral efficiency with ten-times less antennas than previously proposed systems \cite{Marzetta_WC10}. Note that the number of antennas is still quite large considering the current radio frequency (RF) techniques. 

In this paper, we consider a network architecture (named cloud BS) to implement feasible massive MIMO systems. The network massive MIMO system consists of multiple radio units (RUs) connected with one another by optical fibers, and further connected to a centralized digital unit (DU), as illustrated in Figs.~\ref{Fig:SysModel1} and \ref{Fig:SysModel2}.\footnote{To avoid confusion, we use RU instead of base station hereafter.} Through the optical fibers, each RU can share the power, data messages, and channel state information; thus network massive MIMO with relatively small number of antennas can be treated as a single-cell massive MIMO with large-scale antennas. For algorithm designs, we consider ZF and MF precodings with two normalization techniques (vector/matrix normalizations). Most prior work on multiuser MIMO algorithms has paid little attention to this issue but we will show that performances differ according to each normalization technique. In this paper, we analyze i) which precoding normalization method is better as far as precoding techniques and ii) which precoding method is appropriate for cell-boundary users. To the best of our knowledge, such analysis has yet to be done in massive/multiuser MIMO systems.

This paper is organized as follows. In Section \ref{Section2}, we introduce the system model for network massive MIMO systems. We also explain the problem statement in respect to precoding normalization methods and beamforming techniques for cell-boundary users. In Section \ref{Section3}, we analyze i) rate lower and upper bounds, ii) ergodic performance of ZF- and MF-precoding, and iii) network performance for cell-boundary users. Numerical results and discussions are presented in Section \ref{Section4}. Section \ref{Section5} presents our conclusions.

\section{System Model and Problem Statement}\label{Section2}
In this section, we introduce the basic notation used in this paper and the network massive MIMO system model.\footnote{Throughout this paper, we use upper and lower case boldfaces to describe matrix $\pmb{A}$ and vector $\pmb{a}$, respectively. We denote the inverse, transpose and the Hermitian of matrix $\pmb{A}$ by $\pmb{A}^{-1}$, $\pmb{A}^{T}$ and $\pmb{A}^{*}$, respectively. $||\pmb{A}||_{F}$ indicates the Frobenius norm of matrix $\pmb{A}$. The notation of expectation is represented by $\mathbb{E}$.
} 

\subsection{System Model: Network Massive MIMO}
\begin{figure}[t]
 \centerline{\resizebox{0.8\columnwidth}{!}{\includegraphics{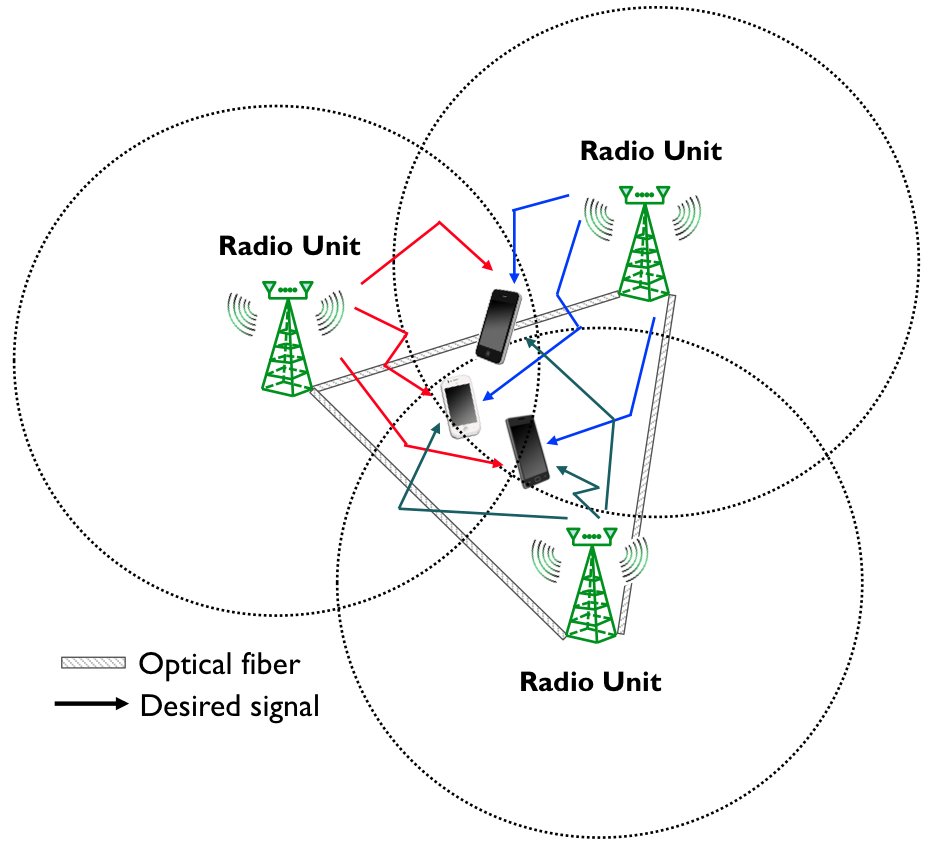}}}
  \caption{System model of network massive MIMO with radio units (RUs) and cell-boundary users.}
  \label{Fig:SysModel1}
\end{figure}

\begin{figure}[t]
 \centerline{\resizebox{0.8\columnwidth}{!}{\includegraphics{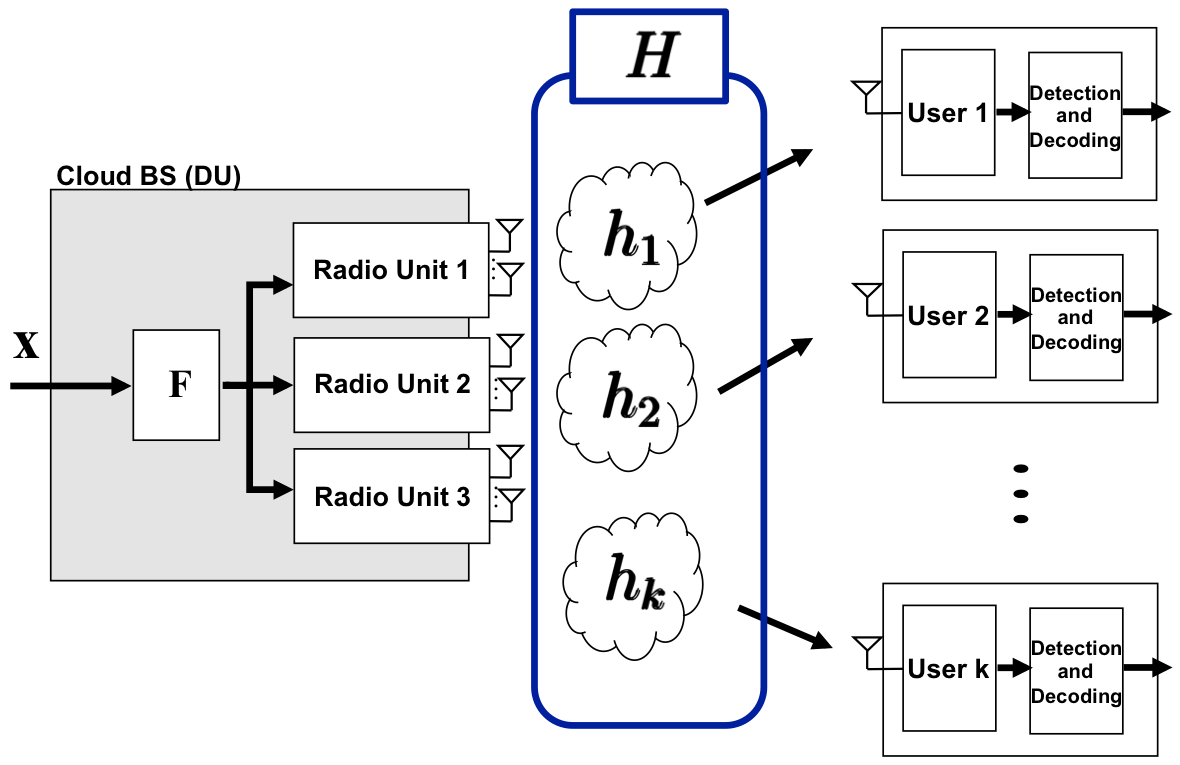}}}
  \caption{Block diagram of network massive MIMO system. One DU consists of three RUs connected by optical fibers. }
  \label{Fig:SysModel2}
\end{figure}

Consider a cooperative network massive MIMO system as shown in Figs. {\ref{Fig:SysModel1}} and {\ref{Fig:SysModel2}}. One DU controls three RUs and $K$ users. Each RU is connected with one another by optical fibers. We assume that each RU has $N_t$ transmit antennas and each user equipment (UE) is equipped with one receive antenna. 
We also assume that the channel is flat fading and the elements of a channel matrix are modeled as independent complex Gaussian random variables with zero mean and unit variance. The channel between cloud BS (one DU and three RUs) and the $k$-th user is denoted by an $1 \times M$ row vector $\pmb{h}_{k}^T$ ($k=1, 2, \cdots, K$), where $M$ represents the number of cloud BS antennas $3N_{t}$. A $K \times M$ channel matrix $\pmb{H}$ between cloud BS and all UEs consists of channel vectors $\pmb{h}_{k}^T$. Let $\pmb{g}_k$ denote the column vector of  transmit precoding and $s_k$ represent the transmit symbol for the $k$-th UE. Also, let $n_k$ be the additive white Gaussian noise vector. Then, the received signal at the $k$-th UE is expressed by
\begin{align}
y_k=\underbrace{\sqrt{P}\pmb{h}_k^T\pmb{g}_ks_k}_{\text{desired signal}}+\underbrace{\sum_{\ell=1,\ell\neq k}^{K}\sqrt{P}\pmb{h}_k^T\pmb{g}_{\ell}s_\ell}_{\text{interference}}+n_k\label{received sig eq}
\end{align}
where $P$ denotes the total network transmit power across three RUs.

\subsection{Problem Statement: Precoding Normalization Perspective}
Eq. (\ref{received sig eq}) contains the desired signal, interference, and noise terms. To eliminate the interference term, to maximize the signal-to-noise-ratio (SNR), we use the following precoding. 
\begin{align}
&\text{ZF} : \pmb{F}=\pmb{H}^*(\pmb{H}\pmb{H}^*)^{-1}=[\pmb{f}_1~ \pmb{f}_2~ \cdots~ \pmb{f}_k]\nonumber\\
&\text{MF} : \pmb{F}=\pmb{H}^*=[\pmb{f}_1~ \pmb{f}_2~ \cdots~ \pmb{f}_k]\nonumber
\end{align}
where $\pmb{F}$ is a precoding matrix consisting of each column vector $\pmb{f}_k$.

To satisfy the power constraint, we need to normalize the precoding matrix. In this paper, as mentioned earlier, two methods, i.e., vector/matrix normalizations, are considered. The normalized transmit beamforming vectors (columns of a precoding matrix) with vector/matrix normalizations are given as $\pmb{g}_k=\pmb{f}_k/(\sqrt{K}||\pmb{f}_k||)$ and $\pmb{g}_k=\pmb{f}_k/||\pmb{F}||_F$, respectively. 


\subsubsection{ZF/MF with vector normalization}
The received signal at the $k$-th UE can be expressed as follows:
\begin{align}
y_k=\sqrt{P}\pmb{h}_k^T\frac{\pmb{f}_k}{\sqrt{K}||\pmb{f}_k||}s_k+\sum_{\ell=1,\ell\neq k}^{K}\sqrt{P}\pmb{h}_k^T\frac{\pmb{f}_{\ell}}{\sqrt{K}||\pmb{f}_{\ell}||}s_\ell+n_k.\label{ZFMF vector norm. eq}
\end{align}

\subsubsection{ZF / MF with matrix normalization}
Similarly, we can rewrite the received signal with matrix normalization as below:
\begin{align}
y_k=\sqrt{P}\pmb{h}_k^T\frac{\pmb{f}_k}{||\pmb{F}||_F}s_k+\sum_{\ell=1,\ell\neq k}^{K}\sqrt{P}\pmb{h}_k^T\frac{\pmb{f}_{\ell}}{||\pmb{F}||_F}s_\ell+n_k.\label{ZFMF matrix norm. eq}
\end{align}

\section{Asymptotic Rate Bounds: ZF and MF cases}\label{Section3}
In this section, we derive the capacity bounds and show which normalization method is suitable for ZF- and MF-type precoding. Based on our analytical results, we will also show which precoding technique is desired for cell-boundary users. 

\subsection{Capacity Bound}
Using \emph{Jensen's Inequality} of convex and concave functions, we can get the capacity's lower and upper bounds as follows:
\begin{align}
\log_2\left(1+\frac{1}{\mathbb{E}\left(\frac{I+N}{S}\right)}\right)&\le\mathbb{E}\left(\log_2\left(1+\frac{S}{I+N}\right)\right)\nonumber\\
&\le \log_2\left(1+\mathbb{E}\left(\frac{S}{I+N}\right)\right).
\end{align}

\subsection{Ergodic performance of ZF precoding}
\subsubsection{Vector normalization-lower bound}
From (\ref{ZFMF vector norm. eq}), we can derive the lower bound of vector normalization in the ZF case as follows:
\begin{align}
\mathbb{E}_{\text{ZF}}\left\{\frac{I+N}{S}\right\}&=\mathbb{E}\left\{\frac{P\sum_{\ell=1,\ell\neq k}^{K}\left|\pmb{h}_k^T\frac{\pmb{f}_{\ell}}{\sqrt{K}||\pmb{f}_{\ell}||}\right|^2+1}
{P\left|\pmb{h}_k^T\frac{\pmb{f}_k}{\sqrt{K}||\pmb{f}_k||}\right|^2}\right\}\nonumber \\
&\overset{(a)}{=}\mathbb{E}\left\{\frac{1}{P\frac{1}{K||\pmb{f}_k||^2}}\right\}
=\mathbb{E}\left\{\frac{K||\pmb{f}_k||^2}{P}\right\}\nonumber \\
&=\frac{1}{P}\mathbb{E}\left\{\frac{K||\pmb{f}_k||^2}{1}\right\}
\overset{(b)}{=}\frac{1}{P}\frac{K}{M-K+1}\label{VectorNorm Low eq}
\end{align}
where $(a)$ follows the property of ZF precoding. By using ZF, the interference term is perfectly eliminated and $\pmb{h}_k^T\pmb{f}_k$ is equal to one in $(a)$. Equality $(b)$ results from the diversity order of ZF, as shown in \cite{Wong_08}.

\subsubsection{Matrix normalization-upper bound}
From (\ref{ZFMF vector norm. eq}), the upper bound of matrix normalization in the ZF case can be expressed as
\begin{align}
\mathbb{E}_{\text{ZF}}\left\{\frac{S}{I+N}\right\}&=\mathbb{E}\left\{\frac{P\left|\pmb{h}_k^T\frac{\pmb{f}_k}{||\pmb{F}||_F}\right|^2}
{P\sum_{\ell=1,\ell\neq k}^{K}\left|\pmb{h}_k^T\frac{\pmb{f}_{\ell}}{||\pmb{F}||_F}\right|^2+1}\right\}\nonumber \\
&=\mathbb{E}\left\{P\left|\frac{1}{||\pmb{F}||_F}\right|^2\right\}
\overset{(c)}{\approx}P\frac{1}{\frac{K}{M-K}}\nonumber\\
&=\frac{P(M-K)}{K}\label{MatrixNorm Upp eq}
\end{align}
where $(c)$ can be obtained by
\begin{align}
\mathbb{E}\{||\pmb{F}||_F^2\}&=\mathbb{E}\left\{\text{tr}(\pmb{H}^*(\pmb{H}\pmb{H}^*)^{-1}(\pmb{H}\pmb{H}^*)^{-1}\pmb{H})\right\}\nonumber \\
&=\mathbb{E}\left\{\text{tr}(\pmb{H}\pmb{H}^*(\pmb{H}\pmb{H}^*)^{-1}(\pmb{H}\pmb{H}^*)^{-1})\right\}\nonumber\\
&=\mathbb{E}\left\{\text{tr}((\pmb{H}\pmb{H}^*)^{-1})\right\}\nonumber \\
&=\frac{K}{M-K}\nonumber
\end{align}
using the property of Wishart matrix of random matrix theory ~\cite{RandomMatrixBook}.

\subsubsection{Performance comparison of ZF}
To find which normalization technique is better in ZF, we compare the lower rate bound of vector normalization $(\mathcal{R}_{\text{ZF}_{\text{vec}}}^L)$ with the upper rate bound of matrix normalization $(\mathcal{R}_{\text{ZF}_{\text{mat}}}^U)$ in ZF; thus the gap is given by
\begin{align}
&~~~\mathcal{R}_{\text{ZF}_{\text{vec}}}^L-\mathcal{R}_{\text{ZF}_{\text{mat}}}^U\nonumber \\
&=\log_2\left(1+\frac{P(M-K+1)}{K}\right)-\log_2\left(1+\frac{P(M-K)}{K}\right)\nonumber\\
&=\log_2\left(\frac{K+P(M-K)+P)}{K+P(M-K)}\right)\nonumber \\
&=\log_2\left(1+\frac{P}{K+P(M-K)}\right)\ge 0. \label{comparisonZF}
\end{align}
From (\ref{comparisonZF}), we could conclude that, in the ZF case, vector normalization is always better than matrix normalization. 


\subsection{Ergodic performance of MF precoding}
\subsubsection{Matrix normalization-lower bound}
From (\ref{ZFMF matrix norm. eq}), the lower rate bound of matrix normalization is given as follows:
\begin{align}
\mathbb{E}_{\text{MF}}\left\{\frac{I+N}{S}\right\}&=\mathbb{E}\left\{\frac{P\sum_{\ell=1,\ell\neq k}^{K}\left|\pmb{h}_k^T\frac{\pmb{f}_{\ell}}{||\pmb{F}||_F}\right|^2+1}
{P\left|\pmb{h}_k^T\frac{\pmb{f}_k}{||\pmb{F}||_F}\right|^2}\right\}\nonumber \\
&\overset{(d)}{\approx}\mathbb{E}\left\{\frac{P(K-1)\frac{M}{MK}+1}{P\frac{M^2+M}{MK}}\right\}
=\frac{\frac{P(K-1)}{K}+1}{\frac{P(M+1)}{K}}\nonumber \\
&=\frac{P(K-1)+K}{P(M+1)}. \label{MatrixNorm Low eq}
\end{align}
Note that MF precoding asymptotically eliminates the interference term in (1) when the cloud BS has reasonably large scale transmit antennas. Applying the properties of random vectors and the law of large numbers to $(d)$, $|\pmb{h}_k^T\pmb{f}_{\ell}|^2$ is $M$, $|\pmb{h}_k^T\pmb{f}_k|^2$ is $M^2+M$, and $||\pmb{F}||_F^2$ can be expressed as $MK$. 


\subsubsection{Vector normalization-upper bound}
In a similar way, we can get the upper rate bound of vector normalization as follows:
\begin{align}
\mathbb{E}_{\text{MF}}\left\{\frac{S}{I+N}\right\}=\mathbb{E}\left\{\frac{P\left|\pmb{h}_k^T\frac{\pmb{f}_k}{\sqrt{K}||\pmb{f}_k||}\right|^2}
{P\sum_{\ell=1,\ell\neq k}^{K}\left|\pmb{h}_k^T\frac{\pmb{f}_{\ell}}{\sqrt{K}||\pmb{f}_{\ell}||}\right|^2+1}\right\}\nonumber \\
\approx\mathbb{E}\left\{\frac{P\frac{(M+1)}{K}}{P\frac{K-1}{K}+1}\right\}
=\frac{P(M+1)}{P(K-1)+K}.\label{VectorNorm Upp eq}
\end{align}

\begin{figure*}[!b]
\hrulefill
\begin{align}
&~~~\mathcal{G}_{\text{MF}_{\text{mat}}}^L-\mathcal{G}_{\text{ZF}_{\text{vec}}}^L\nonumber \\
 &=\frac{4(M+1)P-M^2P^2+(MP+P-1)\sqrt{MP(MP-4)-4P}}{(M+1)(2MP+P+1)}\nonumber \\
 &=\frac{-((MP-2)^2-4(P+1))+(MP+P-1)\sqrt{(MP-2)^2-4(P+1)}}{(M+1)(2MP+P+1)}\label{Diff_gradient}
\end{align}
\end{figure*}

\subsubsection{Performance comparison of MF}
We also compare the rate bounds, and the gap is given by
\begin{align}
&~~~\mathcal{R}_{\text{MF}_{\text{mat}}}^L-\mathcal{R}_{\text{MF}_{\text{vec}}}^U\nonumber \\
&=\log_2\left(1+\frac{P(M+1)}{P(K-1)+K}\right)-\log_2\left(1+\frac{P(M+1)}{P(K-1)+K}\right)\nonumber \\
&=0 \label{comparisonMF}
\end{align}
where $\mathcal{R}_{\text{MF}_{\text{mat}}}^L$ and $\mathcal{R}_{\text{MF}_{\text{vec}}}^U$ denote the lower rate bound of matrix normalization and the upper rate bound of vector normalization, respectively. From (\ref{comparisonMF}), we confirm that, for MF precoding, matrix normalization is always better than vector normalization. 


\begin{figure}[t]
 \centerline{\resizebox{0.85\columnwidth}{!}{\includegraphics{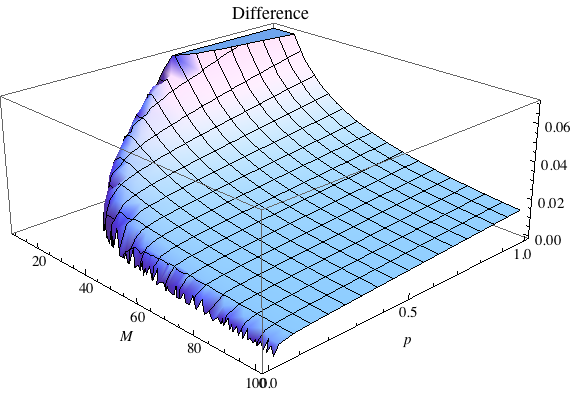}}}
  \caption{The difference of the gradient between ZF and MF at $K_{\text{cross}}$ when $P$ is very small (almost zero) and $M$ is much larger than $P$. The difference is always positive ($>0$).}
  \label{Difference}
\end{figure}

\subsection{Low SNR Analysis}\label{subsection D}
In Section~\ref{Section3}, we investigated which normalization is appropriate for ZF and MF precoding techniques. In this section, we will analyze which precoding technique is better for cell-boundary users, i.e., low signal-to-interference-plus-noise (SINR) users. 


Both $\mathcal{R}_{\text{ZF}_{\text{vec}}}^L$ and $\mathcal{R}_{\text{MF}_{\text{mat}}}^L$ are concave functions. Also, unlike $\mathcal{R}_{\text{ZF}_{\text{vec}}}^L$, $\mathcal{R}_{\text{MF}_{\text{mat}}}^L$ is a monotonic increasing function; thus, two cross points exist: one is when the number of users $K$ is one, the other is when the number of users $K$ has the following value with a large $M$ approximation:\footnote{We derive this by using two rate bounds equations and omit the derivations here.}
\begin{align}
K_{\text{cross}}=\frac{P(M+1)}{1+P}
\label{CrossPoint eq}
\end{align}
where, $K_{\text{cross}}$ denotes the crossing point as functions of the number of cloud BS antennas $(M)$ and the total transmit power $(P)$. In (\ref{CrossPoint eq}), intuitively, $K_{\text{cross}}$ should be greater than zero and integer. Therefore, at the low SNR regime, (\ref{CrossPoint eq}) goes to zero, thus the limit becomes one with a $K_{\text{cross}}$ constraint.
This means that as SNR decreases the cross point shifts to the left. At $K_{\text{cross}}$, we check the difference of the gradient between ZF and MF. If the gradient of ZF is larger than that of MF, the rate of ZF with vector normalization is larger than that of MF when $K \ge K_{\text{cross}}$. In the other case, the rate of MF with matrix normalization is larger than that of ZF when $K \ge K_{\text{cross}}$. The difference of the gradient between ZF and MF is expressed as (\ref{Diff_gradient}) on the bottom,
where $\mathcal{G}_{\text{MF}_{\text{mat}}}^L$ denotes the gradient of the $\mathcal{R}_{\text{MF}_{\text{mat}}}^L$ curve at $K_{\text{cross}}$. Similarly, $\mathcal{G}_{\text{ZF}_{\text{vec}}}^L$ is the gradient of the $\mathcal{R}_{\text{ZF}_{\text{vec}}}^L$ curve at $K_{\text{cross}}$. In general, cell-boundary users have relatively low SINR and, as we assumed, the cloud BS has large-scale antennas, meaning $M$ is much larger than $P$. Therefore, if $K_{\text{cross}}$ exists, (\ref{Diff_gradient}) is always positive. We also confirm this through numerical comparisons as shown in~Fig.~\ref{Difference}. From this observation, we realize that MF precoding is suitable for cell-boundary users if the number of active users is larger than $K_{\text{cross}}$.


\begin{figure}[t]
 \centerline{\resizebox{0.85\columnwidth}{!}{\includegraphics{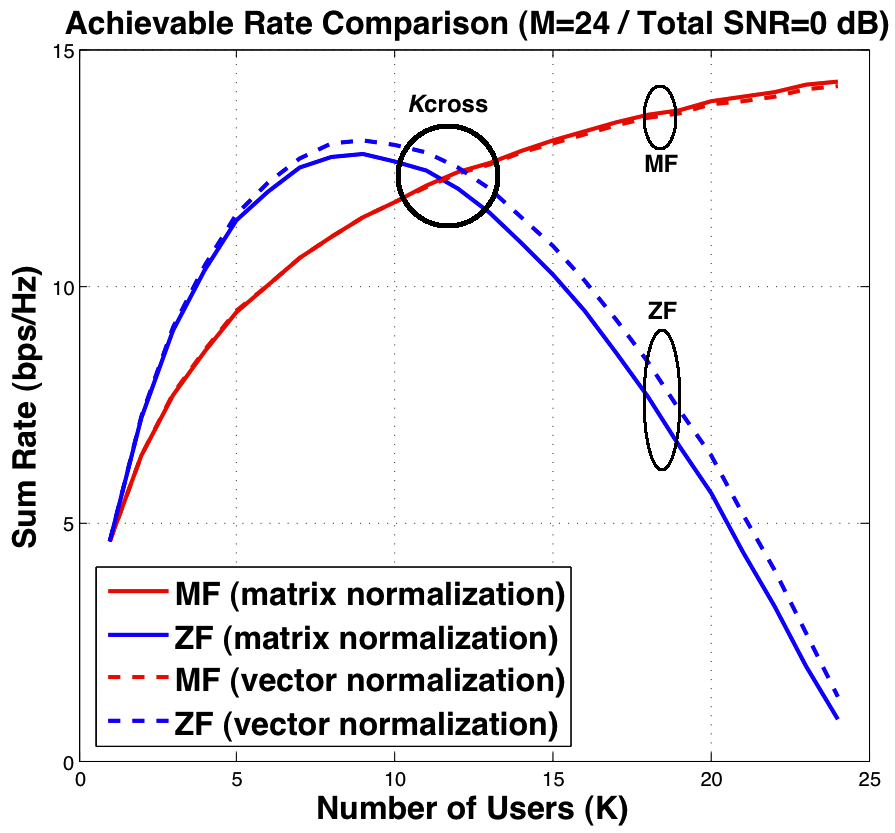}}}
  \caption{Achievable rate vs. the number of cell-boundary users, where $M$ = 24, $K$ = [1, 24], and total SNR = 0 dB.}
  \label{0dB}
\end{figure}

\section{Discussion}\label{Section4}
For numerical comparisons, we assume that each RU has eight transmit antennas; thus the cloud BS has a total of 24 antennas. Note that any number of antennas can be used and this constraint is not really related to our system. This assumption is based on a 3GPP LTE-advanced's parameter; Release 10 supports eight Node B antennas \cite{Baker_LTE12}. Instead of increasing the number of antennas at each transmitter, we propose using the more feasible cloud concept. Indeed, having more than eight transmit antennas would be difficult due to pilot overhead and other system constraints. We are not arguing here that we have to have only eight antennas, rather we show the gain of massive MIMO that can be achieved through a simple cooperation with relatively small number of antennas at each RU.

\begin{figure}[!t]
 \centerline{\resizebox{0.85\columnwidth}{!}{\includegraphics{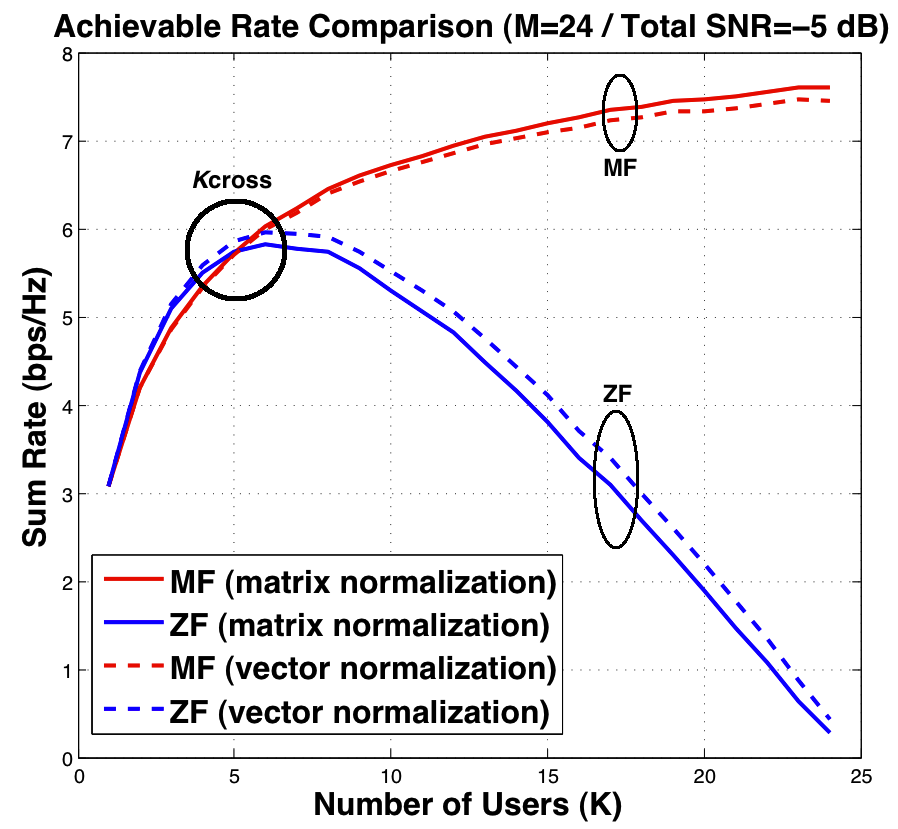}}}
  \caption{Achievable rate vs. the number of cell-boundary users, where $M$ = 24, $K$ = [1, 24], and total SNR = -5 dB.}
  \label{-5dB}
\end{figure}

In Fig. \ref{0dB}, we compare the achievable sum rates of ZF precoding with MF precoding when the total transmit SNR is 0~dB. As was shown in Section \ref{Section3}, ZF with vector normalization is better. In contrast, MF with matrix normalization is better at getting an improved sum rate performance. 
Fig.~\ref{-5dB} illustrates the achievable sum rates of ZF- and MF-precoding with -5 dB transmit SNR. The result is similar to that of Fig. \ref{0dB}. As we mentioned in Section \ref{subsection D}, in the low SNR regime (meaning that users are located at cell-boundary), using MF precoding is generally better when the number of active users is larger than $K_{\text{cross}}$. Also, we realize through Figs.~\ref{0dB} and \ref{-5dB} that as SNR decreases $K_{\text{cross}}$ shifts to the left. The derived $K_{\text{cross}}$ expression is also verified through numerical results. We summarize our conclusions in Tables \ref{Table1} and \ref{Table2}.


\section{Conclusion}\label{Section5}
In this paper, we proposed network massive MIMO systems supporting multiple cell-boundary UEs. For precoding designs, we first derived the achievable rate bounds of zero-forcing (ZF) and matched filter (MF) with vector or matrix normalization. Through analytical and numerical results, we confirmed that vector normalization is better for ZF while matrix normalization is better for MF. We also investigated the optimal mode switching point as functions of the number of active users in a network and the total transmit power. In future work, we will consider limited cooperation among RUs and cooperation delay.  

%
%

\section*{Acknowledgment} The first author would like to thank Yeon-Geun Lim and June Hwang for helpful discussions.

\bibliographystyle{IEEEtran}

\bibliography{GlobeCom2012_references}

\begin{thebibliography}{10}
\providecommand{\url}[1]{#1}
\csname url@samestyle\endcsname
\providecommand{\newblock}{\relax}
\providecommand{\bibinfo}[2]{#2}
\providecommand{\BIBentrySTDinterwordspacing}{\spaceskip=0pt\relax}
\providecommand{\BIBentryALTinterwordstretchfactor}{4}
\providecommand{\BIBentryALTinterwordspacing}{\spaceskip=\fontdimen2\font plus
\BIBentryALTinterwordstretchfactor\fontdimen3\font minus
  \fontdimen4\font\relax}
\providecommand{\BIBforeignlanguage}[2]{{%
\expandafter\ifx\csname l@#1\endcsname\relax
\typeout{** WARNING: IEEEtran.bst: No hyphenation pattern has been}%
\typeout{** loaded for the language `#1'. Using the pattern for}%
\typeout{** the default language instead.}%
\else
\language=\csname l@#1\endcsname
\fi
#2}}
\providecommand{\BIBdecl}{\relax}
\BIBdecl

\bibitem{Chae_SPMag_07}
D.~Gesbert, M.~Kountouris, {R. W. Heath, Jr.}, C.-B. Chae, and T.~Salzer,
  ``Shifting the {MIMO} paradigm: From single user to multiuser
  communications,'' \emph{IEEE Sig. Proc. Mag.}, vol.~24, no.~5, pp. 36--46,
  Oct. 2007.

\bibitem{SpencerSwindle04}
Q.~Spencer, A.~L. Swindlehurst, and M.~Haardt, ``Zero-forcing methods for
  downlink spatial multiplexing in multiuser {MIMO} channels,'' \emph{IEEE
  Trans. Sig. Proc.}, vol.~52, pp. 462--471, Feb. 2004.

\bibitem{Chae_JSAC07}
C.-B. Chae, D.~Mazzarese, N.~Jindal, and {R. W. Heath, Jr.}, ``Coordinated
  beamforming with limited feedback in the {MIMO} broadcast channel,''
  \emph{IEEE Jour. Select. Areas in Comm.}, vol.~26, no.~8, pp. 1505--1515,
  Oct. 2008.

\bibitem{Chae_SPL09}
C.-B. Chae and {R. W. Heath, Jr.}, ``On the optimality of linear multiuser
  {MIMO} beamforming for a two-user two-input multiple-output broadcast
  system,'' \emph{IEEE Sig. Proc. Lett.}, vol.~16, no.~2, pp. 117--120, Feb.
  2009.

\bibitem{Marzetta_06}
T.~L. Marzetta, ``How much training is required for multiuser {MIMO}?'' in
  \emph{Proc. of Asilomar Conf. on Sign., Syst. and Computers}, 2006, pp.
  359--363.

\bibitem{Ngo_TCOM12}
H.~Q. Ngo, E.~G. Larsson, and T.~L. Marzetta, ``Energy and spectral efficiency
  of very large multiuser {MIMO} systems,'' \emph{IEEE Trans. Comm.}, 2012,
  available: http://arxiv.org/abs/1112.3810.

\bibitem{Marzetta_WC10}
T.~L. Marzetta, ``Noncooperative cellular wireless with unlimited numbers of
  base station antennas,'' \emph{IEEE Trans. Wireless Comm.}, vol.~9, no.~11,
  pp. 3590--3600, Nov. 2010.

\bibitem{Chae_NCBF08}
C.-B. Chae, S.~Kim, and {R. W. Heath, Jr.}, ``Network coordinated beamforming
  for cell-boundary users: Linear and non-linear approaches,'' \emph{IEEE Jour.
  Select. Topics in Sig. Proc.}, vol.~3, no.~6, pp. 1094--1105, Dec. 2009.

\bibitem{Chae_IACBF09}
C.-B. Chae, I.~Hwang, {R. W. Heath, Jr.}, and V.~Tarokh, ``Interference
  aware-coordinated beamforming in a multi-cell system,'' \emph{IEEE Trans.
  Wireless Comm.}, vol.~11, no.~10, pp. 1--12, Oct. 2012.

\bibitem{Jubin_WC11}
J.~Jose, A.~Ashikhmin, T.~L. Marzetta, and S.~Vishwanath, ``Pilot contamination
  and precoding in multi-cell {TDD} systems,'' \emph{IEEE Trans. Wireless
  Comm.}, vol.~10, no.~8, pp. 2640--2651, Aug. 2011.

\bibitem{Jakob}
J.~Hoydis, S.~ten Brink, and M.~Debbah, ``Massive {MIMO}: How many antennas do
  we need?'' in \emph{Proc. of Allerton Conf. on Comm. Control and Comp.},
  2011, pp. 545--550.

\bibitem{Hoon11}
H.~Huh, G.~Caire, H.~C. Papadopoulos, and S.~A. Ramprashad, ``Achieving large
  spectral efficiency with {TDD} and not-so-many base-station antennas,'' in
  \emph{Proc. IEEE-APS Conf. on Antennas and Prop. for Wireless Comm.}, 2011,
  pp. 1346--1349.

\bibitem{Wong_08}
K.~K. Wong and Z.~Pan, ``Array gain and diversity order of multiuser {MISO}
  antenna systems,'' \emph{Int. J. Wireless Inf. Networks}, vol. 2008, no.~15,
  pp. 82--89, May 2008.

\bibitem{RandomMatrixBook}
A.~M. Tulino and S.~Verdu, ``Random matrix theory and wireless
  communications,'' \emph{Foundations and Trends in Comm. and Info. Th.},
  vol.~1, no.~1, 2004.

\bibitem{Baker_LTE12}
{S. Sesia}, {I. Toufik}, and {M. Baker}, \emph{{LTE, The UMTS Long Term
  Evolution: From Theory to Practice}}.\hskip 1em plus 0.5em minus 0.4em\relax
  Wiley, 2012.

\end{thebibliography}

\begin{table}[!t]
\caption{Precoding normalization techniques in network massive MIMO systems.}
\begin{center}
\begin{tabular}{|c||c|}
\hline
& Precoding normalization technique\\
\hline \hline
$\text{ZF}$ & Vector normalization $\ge$ Matrix normalization \\
\hline
$\text{MF}$ & Matrix normalization $\ge$ Vector normalization \\
 \hline
\end{tabular}
\end{center}
\label{Table1}
\end{table}

\begin{table}[!t]
\caption{Desired precoding technique in network massive MIMO systems. ZF $(\text{if}~K \leq K_{\text{cross}})$ and MF $(\text{if}~K \geq K_{\text{cross}})$.}
\begin{center}
\begin{tabular}{|c||c|c|}
\hline
& $K_{\text{cross}}$ & Precoding technique \\
\hline \hline
Cell-center & Large & Zero-forcing \\
\hline
Cell-boundary & Small  & Matched filter \\
 \hline
\end{tabular}
\end{center}
\label{Table2}
\end{table}

\end{document}